\documentclass[aps,pra,twocolumn,superscriptaddress]{revtex4-1} 

 \usepackage{cancel} 
\usepackage{amsfonts}
\usepackage{amsmath}
\usepackage{amssymb}
\usepackage{graphicx,xcolor}
\usepackage{float}
\usepackage{hyperref}
\usepackage{subfigure}
\usepackage{dcolumn}
\usepackage{soul}
\usepackage{ulem}
\usepackage{verbatim}

\begin{document}

\title{   $D$-dimensional three-body bound-state problem with zero range interactions}

\author{D. S. Rosa}

\affiliation{Instituto Tecnol\'{o}gico de Aeron\'{a}utica, DCTA,
  12228-900 S\~{a}o Jos\'{e} dos Campos, SP, Brazil}
  
  \author{T. Frederico}

\affiliation{Instituto Tecnol\'{o}gico de Aeron\'{a}utica, DCTA,
  12228-900 S\~{a}o Jos\'{e} dos Campos, SP, Brazil}

\author{G. Krein}

\affiliation{Instituto de F\'isica Te\'orica, Universidade Estadual Paulista,
Rua Dr. Bento Teobaldo Ferraz, 271-Bloco II, 01140-070 S\~ao Paulo, SP, Brazil}

\author{M. T. Yamashita}

\affiliation{Instituto de F\'isica Te\'orica, Universidade Estadual Paulista,
Rua Dr. Bento Teobaldo Ferraz, 271-Bloco II, 01140-070 S\~ao Paulo, SP, Brazil}

\begin{abstract}

We solved analytically the three-body mass-imbalanced problem embedded in $D$ dimensions  
for zero-range resonantly interacting particles. We derived the negative energy eigenstates 
of the  three-body Schr\"odinger equation by imposing the Bethe-Peierls boundary conditions 
in $D$-dimensions for zero-energy two-body bound states. The solution retrieves the 
Efimov-like discrete scaling factor dependence with dimension. The analytical form of the 
mass-imbalanced three-body bound state wave function can be used to probe the effective 
dimension of asymmetric cold atomic traps for Feshbach resonances tuned close to the Efimov 
limit.
\end{abstract}

\maketitle

\section{Introduction} Magnetically tunable Feshbach resonances in ultracold atomic gases open 
up several possibilities to explore few- and many-body physics~\cite{feshbach}. The access 
to the universal regime, therein the scattering length exceeds in magnitude all other length 
scales of the system, was a significant breakthrough in cold atom 
physics~\cite{D1,review1,review2}. Not only the interactions and energies can be 
freely tuned in ultracold atomic traps, but also the geometry of the system. 
The ability of squeezing the shape of the atomic cloud opens new opportunities for 
studies of few-body effects in such engineered systems.

In the context of few-body problems, one intriguing phenomenon is the Efimov 
effect~\cite{efimov1,efimov0,efimov2}. It consists of an infinite series 
of weakly bound three-body states following a universal geometrical scaling law close to the 
two or three-body threshold. Several ultracold atomic experiments have by
now observed the Efimov effect in homo~\cite{homoexp0,homoexp1,homoexp2,homoexp3} and heteronuclear 
systems~\cite{heteexp1,heteexp2,heteexp3}. In dilute gases, weakly bound Efimov trimer states 
mediate inelastic collisions giving rise to a rich spectrum of atom loss resonances as a 
function of the tunable scattering length. The universal aspects of Efimov physics, 
first proposed in the nuclear physics context, appear over an incredible variety of systems 
covering a wide range of physical scales: atomic gases~\cite{efimovatoms}, Bose polarons~\cite{efimovpolarons0,
efimovpolarons}, dipolar molecules~\cite{efimovdipoles} and strongly interacting photons~\cite{efimovphotons}, 
to name a few examples. 

Despite of many advances in theory and experiments, which allow the continuous changing 
of the geometry of the system and the effective dimension of the trap, associating the Efimov 
geometrical scaling with the squeezed system remain an important property yet to be observed. Historically, it is known 
theoretically the determinant role of the dimension to establish the Efimov effect - as predicted 
in the early 1980's~\cite{lim1,lim2}, the Efimov effect exists in three dimensions but is absent 
in two. The possibility to experimentally observe this prediction only appeared after 
the construction of Bose-Einstein condensates in traps with one~\cite{BEC1D} and two~\cite{BEC2D} 
dimensions. This experimental advance brought together technologies which allowed modifying
continuously the geometry of the cloud. 

The independent and continuous change of one spatial dimension of the trap, allied to the careful 
control of the scattering length, could potentially lead to the observation of a change in the 
Efimov geometrical ratio associated with an {\it effective dimension}. The separation of successive peaks in the three-body recombination loss~\cite{homoexp3}, or 
even by measuring directly the binding energies of the trimers~\cite{dorner} are one of the possible observables to probe and study the vicinity of vanishing the Efimov effect. However, in a trapped system the successive ratios between trimer states may not be necessarily the same, but could depend non-trivially on the properties of Feshbach resonances~\cite{chinNP2017}. Therefore, one should be cautious when associating the separation between recombination peaks or Efimov states to the non-integer dimension in squeezed traps.

Even still lacking clear experimental evidence, studies of three-body systems in reduced 
geometries have been the subject of interest in recent years. Different approaches 
were employed to study dimensional effects in three-body systems close to the Efimov limit. 
This limit is achieved when the dimer binding energy vanishes or, equivalently, the scattering 
length is driven to infinity, which is also known as unitary limit. In such studies, 
the system is embedded in fractional dimension~$D$~\cite{D1,D2,D2rosa,D3,D4,D5}, 
in mixed dimensions~\cite{mixed1,mixed2,mixed3}, in which atoms move in different spatial 
dimensions, or it is squeezed to lower dimensions by changing the shape of an external 
potential~\cite{squeeze2,squeeze-1,squeeze0,squeeze1,squeeze,traped}. An approximate relation between the non-integer dimension $D$ used in this work to the squeezing in one direction 
by an external potential was already derived in~\cite{garridoprr}~
$b_{ho}^2/r_{2D}^2=3(D-2)/(3-D)(D-1)$, where 
$b_{ho}$ is the harmonic oscillator
parameter and is represented in units of the rms radius of the three-body system in
two dimensions $r_{2D}$.

In this work, we provide the first analytical solution for the bound-state wave function with 
finite binding energy for the resonant three-body mass-imbalanced problem in $D$-dimensions. The calculation uses the Bethe-Peierls (BP) boundary condition 
approach~\cite{bethe},  for each pair of resonant particles in the three-body system, extended to 
arbitrary dimensions. For $D=3$, this was the method originally used by Efimov to solve the 
three-boson problem leading to the discovery of the geometrical ratio of the binding
energies~\cite{efimov1,efimov2}.

The method adopted here follows closely Efimov's solution in  coordinate-space using 
hyperspherical coordinates, which is now applied to three-distinct particles in $D$ dimensions. 
In this case of a zero-range interaction, each Faddeev component of the wave function is an eigenstate 
of the free Schr\"odinger eigenvalue equation for a given binding energy. The BP boundary conditions 
are imposed on the full wave function, obtained by summing the three Faddeev components, to account 
for the zero-range interaction.

The Efimov scaling parameter is obtained from the solution of a transcendental equation in $D$ dimensions, 
which comes from the Faddeev components of the wave function for three different particles. The Efimov 
parameter appears naturally in each Faddeev component as a direct consequence of a system of homogeneous 
linear equations. For the particular case of two identical bosonic particles and a distinct one, 
we reproduce the previous results of Ref.~\cite{D2rosa} obtained with the momentum space 
representation.

The analytical solution of the eigenvalue equation for the three-body bound state wave function opens 
the possibility of future explorations of different observables, such as, the 
three-body radius~\cite{johnradii} and the momentum densities~\cite{braatendensity,castindensity}, 
uncovering analytically the scaling laws of these quantities with the binding energy and dimension. 
Such scaling laws, in correspondence with limit cycles, evidences the crucial importance of the 
effective dimension on the Efimov physics and points out the direction for experimental 
investigations.

\section{ Bethe-Peierls boundary condition in $D$-dimensions} 
We derive the BP boundary condition considering 
a system of two non-relativistic spinless particles in $D$-dimensions with a short-range s-wave 
interaction. For relative distances beyond a finite range, two particles are non interacting and 
the radial wave function of the pair, working in units of $\hbar=1$, is known to be~\cite{2bodies}: 
\begin{small}
\begin{equation}
R(r) = \sqrt{\frac{\pi}{2p}} \, r^{1-\frac{D}{2}} \, \bigl[ \cot\delta_{D}(p) \, J_{\frac{D}{2}-1}(pr) 
- \, Y_{\frac{D}{2}-1}(pr) \bigr],
\end{equation}
\end{small}
where $p$ is the relative momentum, $J_{D/2-1}$ and $Y_{D/2-1}$ are the Bessel functions 
of the first and second kind, and the $s$-wave phase-shift $\delta_D(p)$ is given in terms of the
scattering length~$a$ as:
\begin{eqnarray}
\cot\delta_{D}(p) = \frac{Y_{\frac{D}{2}-1}(p\, a)}{J_{\frac{D}{2}-1}(p\, a)}.
\end{eqnarray}
The Bethe-Peierls boundary condition at zero-energy for the contact interaction can now be obtained by 
taking the limit to the origin of the logarithmic derivative of the reduced wave 
function  $u(r) = r^{(D-1)/2}\,R(r)$: 
\begin{equation}
\left[\frac{d}{dr} \log u(r) \right]_{r\to 0}\hspace{-.2cm} 
= \left[ \frac{D-1}{2r} - \frac{D-2}{r - r (r/a)^{D-2}}
\right]_{r\to 0} , 
\label{fullboundcon}
\end{equation}
which reproduces the well known results:
\begin{equation}
\left[\frac{d}{dr} \log u(r) \right]_{r\to 0} = 
\begin{cases} 
- \frac{1}{a}, \hspace{0.25cm}\text{for}\hspace{0.25cm} D = 3 \\
\frac{1}{2r} - \frac{1}{\log (r/a)}, \hspace{0.25cm} \text{for} \hspace{0.25cm} D = 2. 
\end{cases} 
\end{equation}
We use Eq.(\ref{fullboundcon}) to obtain the solution of the three-body Schr\"odinger equation in the unitary limit. 

\section{ Three-body mass imbalanced problem}

We consider three different bosons with masses $m_i$, $m_j$, 
$m_k$, and coordinates $\textbf{x}_{i}$, $\textbf{x}_{j}$ and $\textbf{x}_{k}$. One can eliminate 
the center of mass coordinate and describe the system in terms of two relative Jacobi coordinates. 
One can identify three sets of such coordinates:
\begin{equation}
 \mbox{\boldmath$r$}_{i} = \textbf{x}_{j} - \textbf{x}_{k}\quad\text{and}\quad
 \mbox{\boldmath$\rho$}_{i} = \textbf{x}_i - \frac{m_j\textbf{x}_j+m_{k}\textbf{x}_k}{m_j + m_k} \, ,
\end{equation}
where ($i, j, k$) are taken cyclically among ($1,2,3$). One can choose any of such sets of coordinates to 
solve the three-body Schr\"odinger equation. The Faddeev decomposition of the three-body wave function 
$\Psi(\textbf{x}_{1},\textbf{x}_{2},\textbf{x}_{3})$ amounts to writing it as a sum 
of three two-body wave functions: $\Psi(\textbf{x}_{1},\textbf{x}_{2},\textbf{x}_{3}) = 
\psi^{(1)}(\mbox{\boldmath$r$}_1,\mbox{\boldmath$\rho$}_1) + 
\psi^{(2)}(\mbox{\boldmath$r$}_2,\mbox{\boldmath$\rho$}_2)
+ \psi^{(3)}(\mbox{\boldmath$r$}_3,\mbox{\boldmath$\rho$}_3)$, where we  omitted the center 
of mass plane wave. Each component satisfies the free
Schr\"{o}dinger eigenvalue equation:
\begin{equation}\label{eq:schr3B}
\left[\frac{1}{2\eta_{i}}\nabla^{2}_{\mbox{\boldmath$r$}_i} +\frac{1}{2\mu_{i}} 
\nabla^{2}_{\mbox{\boldmath$\rho$}_i}
- E\right] \psi^{(i)} (\mbox{\boldmath$r$}_i,\mbox{\boldmath$\rho$}_i) = 0,
\end{equation}
where  $E$ is the system energy and the reduced masses are given by
 $\eta_{i} = m_{j}m_{k}/(m_{j}+m_{k})$ and $ \mu_{i} = {m_{i}(m_{j}+m_{k})}/({m_{i}+m_{j}+m_{k}}).$
The BP boundary condition applies
to the total wave function; when applied to the chosen coordinates pair 
$(\mbox{\boldmath$r$}_i,\mbox{\boldmath$\rho$}_i)$,
it reads, in the unitary limit $a\rightarrow \infty$: 
\begin{equation}
\label{eq:BP3B}
\hspace{-0.2cm}\left[\frac{\partial}{\partial r_i}  
r_{i}^{\frac{D-1}{2}}\Psi(\mbox{\boldmath$r$}_i,\mbox{\boldmath$\rho$}_i)
\right]_{r_i\rightarrow 0} = \frac{3-D}{2} 
\left[\frac{\Psi(\mbox{\boldmath$r$}_i,\mbox{\boldmath$\rho$}_i)}
{r_{i}^{\frac{3-D}{2}}}\right]_{r_i\rightarrow 0}.
\end{equation}
This solution strategy was applied to different particles and spins in 
Ref.~\cite{BulgacSJNP1975} and we adapt it to $D$-dimensions following 
closely  Efimov's original derivation~\cite{efimov1,efimov2}.

For convenience, we can simplify the form of the kinetic energies by introducing the new 
coordinates:
\begin{equation}
 \mbox{\boldmath$r$}'_{i} = \sqrt{\eta_i}\, \mbox{\boldmath$r$}_{i}\quad\text{and}\quad
 \mbox{\boldmath$\rho$}'_{i} = \sqrt{\mu_i}\, \mbox{\boldmath$\rho$}_{i}\, .
\end{equation}
The three sets of primed coordinates are related to each other by the orthogonal transformations
\begin{eqnarray}
\mbox{\boldmath$r$}'_{j}& =& - \mbox{\boldmath$r$}'_{k}\cos\theta_i + \mbox{\boldmath$\rho$}'_{k}
\sin \theta_i, \nonumber \\
\mbox{\boldmath$\rho$}'_{j}& =& - \mbox{\boldmath$r$}'_{k}\sin\theta_i - 
\mbox{\boldmath$\rho$}'_{k}\cos \theta_i,
\end{eqnarray}
where $\tan \theta_i = \left[m_i M/(m_j\ m_k)\right]^{1/2}$, with $M = m_1 + m_2 + m_3$. For bosons in 
the partial-wave channel with vanishing total angular momentum, 
one can define the reduced Faddeev component as 
\begin{equation} 
\chi^{(i)}_0 (r'_{i}, \rho'_i) = \left( r'_{i} \ 
\rho'_{i}\right)^{\frac{D-1}{2}}\psi^{(i)}(r'_i,\rho'_i).
\label{def-chi0}
\end{equation}
The corresponding Schr\"odinger equation for $\chi^{(i)}_0$ is separable in the hyper-spherical coordinates
$r'_i = R \sin \alpha_i$ and $\rho'_i = R \cos \alpha_i $, so that one can write:
\begin{equation}
\chi^{(i)}_0(R,\alpha_{i})  = C^{(i)} F(R)\,G^{(i)}(\alpha_{i})\,,
\end{equation}
where $R^{2}= r_i^{\prime 2}+\rho_i^{\prime2}$ and $\alpha_i = \arctan(r'_i/\rho'_i)$,
with $F(R)$ and $G^{(i)}(\alpha_{i})$ satisfying the following equations:
\begin{eqnarray}
&&\left[- \frac{\partial^{2}}{\partial R^{2}} + \frac{s_{n}^{2}-1/4}{R^{2}} + \kappa_0^2
\right]\sqrt{R}F(R)=0,
 \label{radialwavefunc}
\\
&&\left[- \frac{\partial^{2}}{\partial \alpha_i^{2}} -s_{n}^{2}+\frac{(D-1)(D-3)}{ \sin^2 2 
\alpha_i}\right] G^{(i)}(\alpha_i)=0, 
\label{Eq:angularD}
\end{eqnarray}
where $-\kappa_0^2 = 2 E$, and $s_n$ is the Efimov parameter, to be determined by the BP boundary 
condition.

The definitions $z = \cos 2\alpha_i$ and $G^{(i)} = (1-z^2)^{1/4} g^{(i)}$ turn
Eq.~\eqref{Eq:angularD} into the form of the associated Legendre differential 
equation~\cite{legendrebook} with the known analytical solutions:
\begin{eqnarray} 
G^{(i)}(\alpha_i) &=& \sqrt{\sin2 \alpha_i}\Big[ P_{s_n/2-1/2}^{D/2-1}\,(\cos2\alpha_i) \nonumber \\
&-& \frac{2}{\pi}\tan\big(\pi(s_{n} -1)/2\big) Q_{s_n/2-1/2}^{D/2-1}\,(\cos2\alpha_i)\Big],\ \ \ \ \
\label{Eq:AngSol}
\end{eqnarray}
where $P_{n}^{m}(x)$ and $Q_{n}^{m}(x)$ are the associated Legendre functions. We have imposed the boundary condition that guarantees a 
finite value for the Faddeev component $\psi^{(i)}$ at $\rho_i =0$, which leads the reduced wave function 
to satisfy  $\chi^{(i)}_0(r'_i,\rho'_i=0)=0$. In terms of the hyper-spherical coordinates, it leads to $G^{(i)}(\alpha_i=\pi/2)= 0$, since $\rho_i' = R \cos{\alpha_i}$.

\begin{widetext}
 Therefore, the 
solution for $\psi^{(i)}(r'_i,\rho'_i)$
is given by:
\begin{eqnarray}
\psi^{(i)}(r'_i,\rho'_i) &=&C^{(i)}   \frac{ K_{ s_n}\left(\kappa_0 \sqrt{  r'^{2}_{i}+  \rho'^{2}_{i} } 
\right) }
{ \big(  r'^{2}_{i}+ \rho'^{2}_{i} \big)^{D/2-1/2}}\frac{\sqrt{\sin\big(2 \arctan\left( 
r'_i/\rho'_i\right)\big)}}{\big[\cos\big( \arctan\left( r'_i/\rho'_i\right)\big)\ \sin\big( \arctan\left( 
r'_i/\rho'_i\right)\big)\big]^{D/2-1/2}}
\nonumber \\
&\times&\left[ P_{s_n/2-1/2}^{D/2-1}\Big(\cos\big(2 \arctan( 
r'_i/\rho'_i)\big)\Big)-\frac{2}{\pi}\tan\big(\pi(s_n-1)/2 \big) Q_{s_n/2-1/2}^{D/2-1}\Big(\cos\big(2 
\arctan( r'_i/\rho'_i)\big)\Big)\right]\, ,
\label{wavefunction}
\end{eqnarray}
where $K_{ s_n}$ is the modified Bessel function of the second kind.

One obtains the Efimov parameter  $s_n$ by 
considering that all three pairs of particles are resonant. Then, 
the BP boundary condition, Eq.~\eqref{eq:BP3B}, should be satisfied by the three-body wave function when 
each  relative distance between two of the particles tends to zero, namely $r_i=R \sin\alpha_i\to 0$, 
implying that $\alpha_i\to 0$ for finite hyper-radius $R$. The hyper-radial part of the wave function 
factorizes in the BP boundary condition for each $r_i$, which depends only on the  hyper-angular part of 
each Faddeev component~\eqref{Eq:AngSol}. The resulting homogeneous linear system for the 
coefficients~$C^{(i)}$ reads:
\begin{equation} 
\frac{C^{(i)}}{2}\left[ \left(
\cot\alpha_i\right)^{\!\frac{D-1}{2}} \left( \sin2\alpha_i \frac{\partial}{\partial \alpha_i}  
+ D-3\right) G^{(i)} (\alpha_i) \right]_{\alpha_i\rightarrow 0}
+ (D-2) \left[ 
\frac{C^{(j)} \, G^{(j)}(\theta_k)}
{\left(\sin\theta_k \cos\theta_k\right)^{\!\frac{D-1}{2}}} 
+ \frac{C^{(k)} \, G^{(k)}(\theta_j)}
{\left(\sin\theta_j \cos\theta_j\right)^{\!\frac{D-1}{2}}} 
\right] = 0,
\label{BPsystem}
\end{equation} 
for $i\neq j \neq k$. Taking the three cyclic permutations of $\{i,j,k\}$ one has a
homogeneous system of three linear equations, from which one obtains the Efimov 
parameter $s_n$  by solving the characteristic transcendental equation. 
\end{widetext}

We remark that the key point of this work is the analytical solution, for finite energies, of each Faddeev component for  bound-state systems of the three-distinct particles - this situation is more complex than our previous work given in Ref.~\cite{D2rosa}. The use of the BP boundary condition results in Eq.~\eqref{wavefunction} and, in order to fully define the wave function, Eq.~\eqref{BPsystem} should be solved to determine the Efimov parameter $s_n$ and the relative weights $C^{(i)}$ of the Faddeev components of the wave function.

In the case of a purely imaginary value for the $s_n$   parameter,
the effective potential in  Eq.~(\ref{radialwavefunc}) is attractive, giving rise 
to the well known pathological $1/R^{2}$ interaction. This potential admits a solution 
at any energy with a spectrum ``unbounded from below'' - a phenomenon discovered long ago by 
Thomas~\cite{thomas} and referred to as the ``Thomas collapse''. In particular, 
the transcendental equation  in three-dimensions reduces to the Efimov's one for identical 
bosons~\cite{efimov0}, and in the general case of different particles to the one derived by Bulgac and 
Efimov~\cite{efimov2,BulgacSJNP1975}, 
when the spin is neglected. In these cases the wave function~\eqref{wavefunction} presents the 
characteristic log-periodicity.

The most favourable conditions for the existence of Efimov-like log-periodic solutions 
happen for spinless particles with zero-energy two-body bound state with zero angular 
momentum. The particular case where only two pairs interact resonantly is easily implemented, 
being necessary only to drop one of the equations in Eq.~\eqref{BPsystem} 
and set to zero the Faddeev component corresponding to the non-resonant pair. 
This method may be used also for particles with spin.

\section{ Results and Discussions} 
The present method applies to the bound state of three distinct particles for dimensions $D>2$, where the homogeneous linear system in Eq.~(\ref{BPsystem}) admits nontrivial solutions with purely imaginary values
$s_n \rightarrow is_0$, i.e. in the Efimov region. That happens only for a given range of dimensions 
$2< D < 4$ constrained by  the condition that $s_0(D)\to0$, which also  depends on the mass imbalance in the system.
Here we discuss  some novel examples of triatomic 
systems composed by $^6$Li, $^{23}$Na, $^{87}$Rb and $^{133}$Cs   in $D$-dimensions from the solution of
Eq.~\eqref{BPsystem}, which besides $s_n$ provides the relative weights $C^{(i)}$ of the Faddeev components,  Eq.~\eqref{wavefunction}, and allows to obtain  the configuration space wave function that will be explored in one example in what follows.

In Table~\ref{tab:1} for 
some choices of mass imbalanced systems, we 
show the range of~$D$ values, $D_c^< < D < D_c^>$, and the critical value of the trap parameter for which the 
Efimov effect is present. The results in the table reveal that as the mass 
imbalance increases to heavy-heavy-light, 
the  range of $D$ values for the existence of the  
Efimov effect widens. For two infinitely heavy  
masses, the lowest critical dimension tends to   $D = 2$ 
from above, i.e. $D_c^ < \to 2_+$, while the trap parameter tends to zero. The maximum  critical dimension in that case tends to $D = 4$ from  below, i.e. $D_c^ > \to 4_-$, and the trap length parameter to infinity. Such subtle behaviour is 
clearly seen in  Table~\ref{tab:1} following the pattern from  $^6$Li-$^{23}$Na$_2$ to 
$^6$Li-$^{133}$Cs$_2$ passing through a fully mass imbalanced system. When one of the 
$^{133}$Cs is substituted by a lighter atom, as in $^6$Li-$^{23}$Na-$^{133}$Cs, the region for 
Efimov states shrinks. We observe that among the examples we have discussed, the smallest range of dimensions for the existence of the Efimov effect is found for three identical atoms. In this case
the squeezing length in units of the rms radius in two dimensions is given by $b_{ho}/r_{2D}=0.988$, in this trap configuration the Efimov effect vanish for three identical atoms.

\begin{center}
\begin{figure}[!htb]
\includegraphics[width=8.55cm]{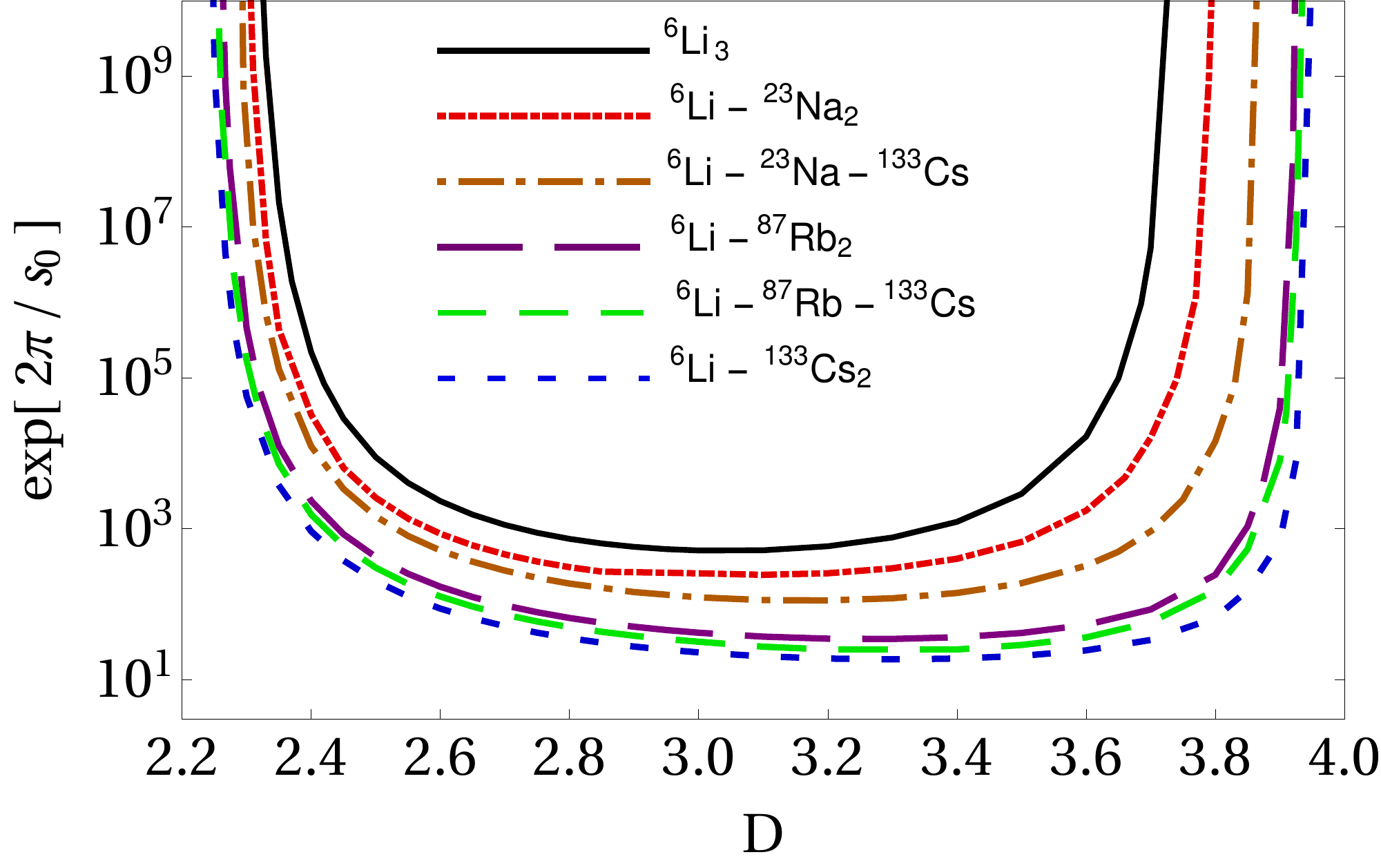}
\caption{Efimov scale parameter as a function of the effective dimension for several mass imbalanced 
system's configuration. }
\label{fig1}
\end{figure}
\end{center}

Figure.~\ref{fig1} displays the geometrical ratio between two successive 
Efimov states as a function of  $D$ for the systems given in Table~\ref{tab:1}.  Noteworthy in the figure, the ratio  of the energies 
of two successive Efimov states varies up to $+\infty$, while large mass asymmetries favor ratios smaller  than those for $D=3$.

\begin{table}[tb]
\centering
\caption{Range of $D$,  $D_c^< < D < D_c^>$, and critical values of the trap parameter allowing  Efimov states for some examples of  mass imbalanced systems.}
\begin{ruledtabular}
\begin{tabular}{lccc}
System   &$b^{<}_{ho}/r_{2D}$   &  $\quad D_c^< \quad$ & $\quad D_c^>\quad$ \\
\hline \hline  \\
$^6$Li$_3$         & 0.988   & 2.297 & 3.755\\[0.2true cm]
$^6$Li\;-\;$^{23}$Na$_2$ &   0.959  & 2.282 & 3.814\\[0.2true cm]
$^6$Li\;-\;$^{23}$Na\;-\;$^{133}$Cs~~~&  0.896    & 2.251 & 3.852\\[0.2true cm]
$^6$Li\;-\;$^{87}$Rb$_2$   &  0.882   & 2.244 & 3.929 \\[0.2true cm] 
$^6$Li\;-\;$^{87}$Rb\;-\;$^{133}$Cs   &  0.864   & 2.235 & 3.938 
\\[0.2true cm]
$^6$Li\;-\;$^{133}$Cs$_2$  & 0.856   & 2.231 & 3.954 \\
\end{tabular}
\end{ruledtabular}
\label{tab:1}
\end{table}

\begin{center}
\begin{figure}[!htb]
\includegraphics[width=8.4cm]{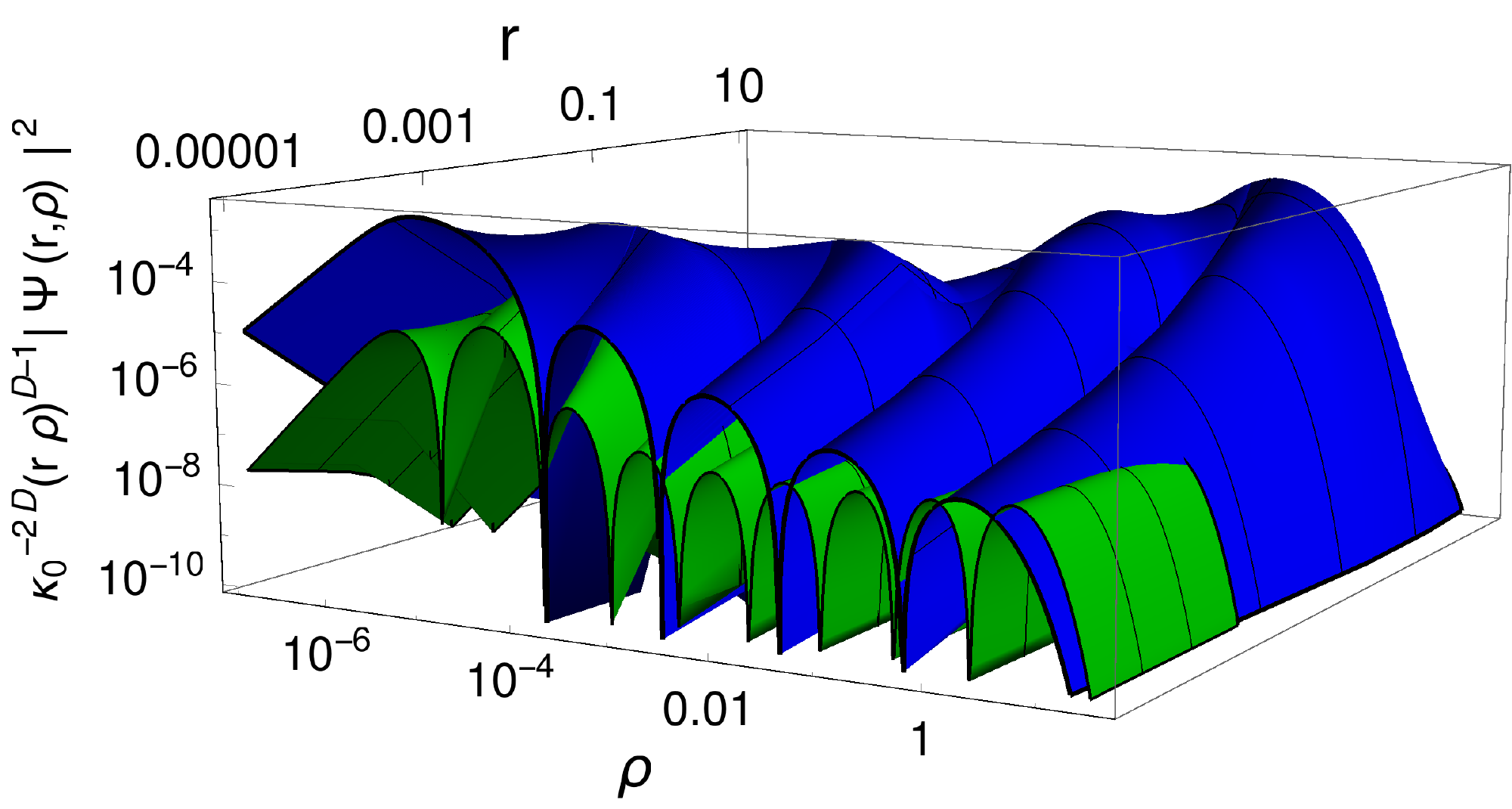}
\vspace{-.5cm}
\caption{Dimensionless radial distribution as a function of dimensionless quantities $r=\kappa_0 r_3$ ($^{133}$Cs-$^{87}$Rb  relative distance) 
and $\rho=\kappa_0 \rho_3$ ($^6$Li relative distance to the $^{133}$Cs-$^{87}$Rb system). We consider the 
three-body system $^{6}$Li-$^{133}$Cs-$^{87}$Rb
for $D=2.5$ (blue) with $b_{ho}/r_{2D}=\sqrt{2}$, and $D=3.0$ (green). The angle between $\vec r$ and $\vec \rho$ is fixed to $\pi/3$.}
\label{fig2}
\end{figure}
\end{center}

 Figure~\ref{fig2} displays the radial distribution of the 
 $^{6}$Li-$^{133}$Cs-$^{87}$Rb molecule  for  $D = 2.5$ and $D = 3.0$, represented respectively by the blue and green surfaces. The
physical realization of $D=2.5$ corresponds to a squeezed trap with $b_{ho}/r_{2D}=\sqrt{2}$, having ratio between
energies of successive shallowest states at unitarity given by $302.5$ ($s_0=1.1$). We recall that for $D=3$, $s_0=1.818$ and  $31.7$ for the energy ratio.
It is possible to observe the log-periodic behavior, fingerprint of an Efimov-like state. The nodes 
of the wave function in the $\rho$ coordinate are located at $\rho_{n+1}\sim{\text e}^{\pi/s_0}\rho_{n}$, and, as expected,  the location depends on $D$~\cite{D2rosa}. 
The oscillations in the $r$~direction, although not visible in the figure,  
are present due to the log-periodicity of the Faddeev component of the wave function coming from 
$K_{\imath s_0}(\kappa_0\sqrt{r_i^{\prime2}+\rho_i^{\prime 2}})$  when 
$\kappa_0\sqrt{r_i^{\prime2}+\rho_i^{\prime 2}}$ attains small enough values.

\section{ Summary}

We presented an analytical solution of the mass-imbalanced three-body problem 
in $D$ dimensions in the Efimov limit. Use of the Bethe-Peierls boundary condition allowed 
us to formulate this problem and in particular show how to compute the Efimov parameter for a wide range 
of mass ratios and dimensions. %
The importance to have a  relatively simple,  analytical way to compute 
the wave function 
for a finite three-body energy opens up the possibility to probe the Efimov physics in ultracold 
atomic systems through radiofrequency spectroscopy~\cite{braatendensity}. Such a technique has 
been used in Ref.~\cite{Wild} to measure Tan's contact parameters~\cite{Tan}, which can be associated 
with thermodynamic properties of the system. Quite recently, the two-body contact was measured 
across the superfluid transition of a planar Bose gas~\cite{Zou}.

Within the perspective of our work, two and three-body contacts can be computed for mass-imbalanced 
systems in $D$-dimensions using Eq.~\eqref{radialwavefunc} by generalizing other known
techniques~\cite{castindensity}, which were applied to three identical bosons in three dimensions. 
The contacts will also allow to address the intriguing phenomenon present in the crossover of the 
discrete and continuum scale symmetry  by decreasing the effective dimension. Then, the system 
evolves from $D=3$ to $D=2$, for which the Efimov effect disappears  - 
in this transition, the log-periodic wave function gives place to a power-law behavior. Such an exciting possibility suggests that the realization of an atomic analogous of unnuclear systems~\cite{unnuclear}, namely unatomic states, may occur in cold traps squeezed from three to two dimensions. We leave the study of such a possibility for a future work.

\section{ ACKNOWLEDGMENTS}

This work was partially supported by: Funda\c{c}\~ao de Amparo \`a Pesquisa do 
Estado de S\~ao Paulo (FAPESP) (grant nos. 2019/00153-8 (M.T.Y.), 2017/05660-0 (T.F.), 
2020/00560-0 (D.S.R.), 2018/25225-9 (G.K.)); Conselho Nacional de Desenvolvimento 
Cient\'{i}fico e Tecnol\'{o}gico (CNPq) (grant nos. 308486/2015-3 (T.F.), 
303579/2019-6 (M.T.Y.), and 309262/2019-4 (G.K.)).

\end{document}